\documentclass[12pt]{iopart}
\usepackage{graphicx}
\usepackage{xcolor}
\usepackage{cite}

\usepackage[pagewise]{lineno}
\begin{document}



\title{An analytical approximation to measure the extinction cross-section using: Localized Waves }

\author{Irving Rond\'on}

\address{ School of Computational Sciences, Korea Institute for Advanced Study, \\
Seoul  0245, Republic of Korea.}

\ead{irondon@kias.re.kr}
\vspace{10pt}

\begin{abstract}
We present a general expression for the optical theorem in terms of Localized Waves.
This representation is well-known and commonly used to generate Frozen waves,  
Xwaves, and other propagation invariant beams.
We analyze several examples using different input beam sources on a circular detector 
to measure the extinction cross-section.
\end{abstract}

\section{Introduction}
Recently, considerable interest has increased to understanding the transversal and longitudinal energy distribution in optics and acoustic waves. In this direction, many applications in several scenarios such as microscopy, material manipulation,  spectroscopy, telecommunication, holography, quantum mechanics, Etc.
Interesting  reviews have been reported in order to show the recent development in engineering applications, and promising opportunities 
using multimode light shaping \cite{Picardo}, and structured beams \cite{Halina,Oleg}.
\\
\\
Due to this demand of different experiments in optics and acoustic \cite{Picardo,Halina,Oleg}, it is essential from the physical point of view to have analytical expressions able to measure the energy distribution (intensity) in a  particular experiment. Generally speaking, this measure can be done using the optical theorem (OT)  \cite{RNewton} .
Essentially, the OT states that the rate at which energy is extinguished due to scattering and absorption at the scattered object \cite{Jackson}. It can help to understand the mechanics and how the absorption and the scattering effects wave propagation throughout a medium and to obtain meaningful features of the object such as the characterization and its physical properties\cite{Bohren}.  
\\
\\
In the last few years, many interesting problems  using beam shape methods  related to optical manipulation and applications of the Lorenz-Mie  theory \cite{Gouesbet1,Gouesbet2}, and  several interesting papers have been  published to generalize the optical theorem in electromagnetic theory and acoustic 
\cite{Zhang2013,Zhang2019,IRondonMPLB,IRondon,Berg2008,Berg2008a,Carney1, Carney2,Marengo1,Marengo2,Marengo3,Mitri1,Mitri2}, in  
quantum mechanics \cite{Gouesbet2009}, seismologic waves \cite{Kees}.  
Also, recently the vector extinction effect in the OT  using radially polarized cylindrical beams  \cite{Krasavin}. 
\\
\\
Technically the OT  for arbitrary can be reviewed for plane waves \cite{Mishchenko1,Mishchenko2}, or using a general representation for structured beams such as the Whitaker integral \cite{Whitaker}.
Using this representation directly to obtain the OT makes a challenging task to solve the extinction integrals analytically and numerically, even for isotropic scattering, and only   numerical results can be applied, which  usually brings some inconveniences for  further analysis, such as  uncertainties concerning the fast oscillating field components
\cite{Mishchenko3}.
\\
\\
However, so far, to our knowledge, there is a gap to describe the OT  using Localized Waves (LW) \cite{Recami}, this is the principal motivation and contribution of this work. 
In this letter, we present for the first time a  general derivation for the optical theorem (OT)  for arbitrary beams using LW. To avoid redundancy about these propagating waves, we refer the reader for a  complete list of works and references in these books \cite{Hugo}.
\\
\\
Using the LW representation is possible to generate several structured beams \cite{Hugo}, just by designing a specific longitudinal spectrum $S(k_z)$ under the proper physical conditions, which can allow producing the transversal optical (or acoustical) intensity shape  \cite{Zamboni}.
\\
Considerable importance in the last few years using LW have been reported in different scenarios such as imaging, microscopy, remote sensing, and optical manipulation  \cite{RABSuarez,Mendoza}. One good example of LW is the so-called "Frozen Waves" (FW) \cite{FW}. 
This beam has unique characteristics and is suited to many applications, such as optical tweezers, remote sensing, atom guiding \cite{Hugo,Zamboni,FW}. For example, 
recently in \cite{Joel} the authors have demonstrated the propagation arrays of "Frozen Waves experimentally" in free space, using synthetic optical holography on a spatial light modulator. 
\\
\\
Our analysis considers the standard derivation approach for the OT for plane waves in the far-field approximation but using LW.
We analyze numerically our results reported by \cite{Markel} using plane waves to, alternatively, measure the power extinguished by a single particle in terms of physical energy fluxes for collimated beam    \cite{Mishchenko1,Mishchenko2,Mishchenko3}, here for structured beams.
We validate our analysis using numerical simulations for two different illumination structured beam probes upon the particle and explore the scattering measured on a  circular detector.

\section{The optical theorem for arbitrary beam in the far field approximation}
\noindent
\label{sec:scpw}
The optical theorem can be derived when a plane wave along a positive $z$ axis is incident on an object \cite{RNewton},  using the approximation that the wave amplitude is larger away from the scattered object. The localized particle is in the origin of the coordinates system and embedded in an infinite homogeneous, and nonabsorbing medium  \cite{Jackson,Bohren}. \\
For simplicity, the time-harmonic factor dependence $e^{-i \omega t}$ is omitted.
\\
In general, the total scalar field is given by
\begin{equation}
	\label{eq1}
	\psi(r)  \approx  \varphi(r) + f\frac{e^{ik r}}{r},
\end{equation}
where the first term represents the incident field and the second is related to the scattered field. $k$ is the wave vector, and
$f$ is the scattering amplitude function (angular dependence). In this study $f$ is considered constant such as in \cite{Mishchenko1,Mishchenko2,Mishchenko3}. In the far-field, $r$ can be approximated as $ r \approx z+ (x^2+y^2)/2R$, where $R$ is the distance from the observation point to the scattered object. Note that for this condition, when $\varphi(r) $ is a plane wave, this expression must satisfy the plane radiation condition in the far-field \cite{RNewton, Jackson,Bohren}.
\\
Using this equation the total intensity is  $ I \approx \psi(r) \psi^*(r) \approx \vert \psi(r) \vert^2 $.
\begin{equation}
	\label{eq2}
	I \approx   \int_{S} \varphi ^* \varphi \,\, da  +  \frac{2 }{R}
	\mathbf{Im} \left( \int _{S}    f^* \varphi \, e^{ i k(x^2+y^2)/2R} \right) da.
\end{equation}
In this expression, the first integral refers to the quantity of energy that the particle receives. The second one, in what follows, is related to the optical theorem or the extinction process. This integration is in the $XY$ plane using the far-field approximation.\\
If $\varphi(r) = e^{ikz}$ is a plane wave, the intensity is 
\begin{equation}
	\label{eq3}
	I\approx A - \frac{4\pi}{k}\,\mathbf{Im}[f],
\end{equation}
where $A$ is the transversal area, the second term is the extinction as a function of the scattering amplitude. 
\\
An interesting question is what happens in this approximation the input scalar field (incident wave) is a structured beam. 
In general, the most general representation to describe any arbitrary incident beam is  in terms of Whittaker’s integral
\cite{Whitaker}.
\begin{equation}
	\label{eq4}	
	\varphi(r)= e^{i k_z z} \int_{-\pi}^{\pi} A(\phi ) e^{ ik_t (x\cos \phi +y \sin \phi )}	
	d\phi,
\end{equation}
where $A(\theta)$  is the angular spectrum of the field, the transverse and longitudinal wave-vector components satisfy the relation $k^2=k_t^2 +k_z^2$ where $k$ is the wave vector.
Note that the Whittaker’s integral can be analytically expressed only for particular functions $A(\phi)$. For example, higher-order Bessel beams are defined by $A(\phi) = e^{i m \phi}$, where $m$ is the azimuthal order of the beam. Mathieu beams are defined by $A(\phi) = C(m, q, \phi) + iS(m, q, \phi)$ are the Mathieu cosine and sine functions. Weber beams are defined by the angular spectrum given by  $A(a,\phi)=\frac{1}{2(\pi\vert \sin \phi \vert^{1/2} )}e^{i a \ln \vert \tan \phi/2 \vert}$, a pedagogical review for invariant beams  see \cite{Uri} and references therein.
\\ 
\\
Substituting these different input invariant beams,  represented by Eq. \ref{eq4} after substitution into  Eq. \ref{eq2}, 
we found that it is not easy to analytically solve (the second term), even for isotropic scattering ($f$ constant). Only numerical results can be applied. Solving  Eq.  \ref{eq2}   is one of the primary and essential contributions reported in this work.
In order to compare and validate our results, we adapted the approximation  using plane waves reported   in Ref. \cite{Mishchenko3}.
\noindent
\\
\\
For this purpose, we explore other general mathematical representations of generating structured beams, and we found that the Localized Waves is the best one to fulfill the physical solution. 
 \\
In this problem, an arbitrary incident LW  with cylindrical symmetry impinging on a circular detector. Following the same physical consideration as in Refs. \cite{Markel,Mishchenko1,Mishchenko2,Mishchenko3}, where the surface is perfectly flat and coincides with the $z=R$ plane, centered in the direction of the incidence wave, with the physical condition $ D/2 <<R$, where $D$ is the detector diameter, and $R$ is the distant point from the scattered object to the detector surface.  A straightforward theoretical setup for this configuration is in \cite{Mishchenko3}.

\section{Basic theoretical background  }
\noindent
In order to solve the problem given by Eq. \ref{eq2}, using Localized Waves (LW),  solutions to the scalar wave equation with cylindrical symmetry,  as a physical superposition of zero-order Bessel beams  \cite{Recami,Hugo}.
\begin{eqnarray}
\label{LongInvB}
\varphi(\rho,z)= \varphi_0    \int_{-\omega/c}^{\omega/c} J_0\left( \rho \sqrt{ \omega^2/c^2-k_z^2}\right) S(k_z) \exp{ (ik_z z )}dk_z,
\end{eqnarray}

where $\varphi_0$  is a certain  amplitude, $\rho = \sqrt{x^2 + y^2}$  is expressed in cylindrical coordinates.  $J_0(\xi)$ is a  zero-order Bessel beam, $S(k_z)$ is a weight function (the longitudinal spectrum) and it is defined over $-\omega/c < k_z < \omega/c$, for non evanescent waves. 
After selecting a longitudinal physical spectrum $S(k_z)$, it is possible to generate different invariant beams with axial symmetry from the physical point of view. It is important to note that Eq. \ref{LongInvB}  satisfies the Maxwell's equations  \cite{Recami,Hugo}. The spectral function $S(k_z)$ can be continuous or discrete \cite{Zamboni}. Using LW solutions allows generating paraxial and nonparaxial beams via the longitudinal wavenumber $k_z=\omega/c$. For paraxial beams, the bandwidth is $\Delta k_z<<\omega/c$ while nonparaxial beams have a wide spectrum bandwidth or the same order of $k_z$. In \cite{Roger}  the authors reported a good description using Maxwell's equations for nonparaxial beams with axial symmetry. This work focuses on nonparaxial beams using a discrete case representation for the longitudinal spectrum to measure the optical cross-section.   
\\
In the following illustrative sections, we briefly describe how to use the LW, given by  Eq. \ref{LongInvB}. This work put forward this representation and our method to measure the optical(acoustical) cross-section for any arbitrary structured field.  

\subsection{ Longitudinal spectrum representation  for  Localized wave solution }
\noindent
Different LW  wave solutions for Eq. \ref{LongInvB} can be obtained 
for a discrete and continuous functions $S(k_z)$.  It is possible to obtain
a generalized solution for the longitudinal spectrum  $S(k_z)$ using the Fourier series  \cite{Goodman}.  For additional technical details, see \cite{Roger}. The discrete case is given by
\begin{equation}
S(k_z) = \sum_{n=-\infty}^{\infty} a_n \exp{\left(i\frac{2\pi n}{\Lambda}k_z \right)},	
\end{equation}
where the coefficient $a_n$ can be  obtained as follow
\begin{equation}
\label{Eq_an}	
	a_n= \frac{1}{\Lambda}  \int_{-\omega/c}^{\omega/c} S(k_z) \exp{ \left( -i\frac{2\pi n}{\Lambda}k_z  \right) } dk_z,
\end{equation}
where $\Lambda=2\omega/c$. For a  continuum  case see \cite{Joel,Zamboni}. 

\subsection{Example 1, LW: using  a constant longitudinal spectrum}
\noindent
The  localized wave solution  for a constant spectrum  is given by
\begin{equation}
	\label{Eq9_Sz0}
	\varphi(\rho,z)=   {\rm sinc} \left[  \sqrt{ \frac{\omega^2}{c^2}   \rho^2 +  \frac{\omega^2}{c^2} z^2  }\right],
\end{equation}
where $S(k_z) = c/2\omega$ \cite{Roger} was solved using  \cite{Abra,Gradshteyn}.

\subsection{Example 2, LW: using a longitudinal Gaussian spectrum}
\noindent
For a longitudinal Gaussian spectrum taking 
$	S(k_z)= \frac{b}{\sqrt{\pi}} \exp{\left[ - b^2(k_z - \bar{k}_z)^2 \right]}$, where $\bar{k}_z$ is related with the central position for the spectrum and $b$ is the amplitude \cite{Roger}. After substituting  Eq. \ref{Eq_an}  into  the  Fourier coefficients formula, we obtain  that the coefficient are given by $a_n\approx \exp{\left( -i\frac{2\pi}{\Lambda}n \bar{k}_z  -\frac{\pi^2 n^2}{\Lambda^2 b^2} \right)}$ in the interval $0\leq k_z \leq \omega/c$. 
Under these considerations the equation \ref{LongInvB} is given by
\begin{equation}
	\label{Eq_10aGSinc}
	\varphi(\rho,z) \approx  \sum_{n=0}^{\infty}  
	\exp{ \left( -i\frac{2\pi}{\Lambda}n \bar{k}_z  -\frac{\pi^2 n^2}{\Lambda^2 b^2}  \right) }  \,\, {\rm sinc} \left[  \sqrt{ \frac{\omega^2}{c^2}  \rho^2 + \left( \frac{\omega}{c} z + \pi n \right)^2  }\right],
\end{equation}
when $n=0$, we recover the former constant longitudinal spectrum solution. Note that the equation Eq. \ref{Eq9_Sz0} and Eq. \ref{Eq_10aGSinc}  satisfies the Maxwell's equations.
For instance, it is possible to study the numerical conditions, convergence, and explanation used to validate this approximation following  Refs.  \cite{Hugo,Roger}.
\\
We obtain different scalar beams shape modulated by $a_n$ and the band limited $sinc$  function, called the "sampling function"  commonly used in digital signal processing and information theory \cite{Goodman}. 
This function is related to the spherical Bessel function $j_n(x)$ of the first kind.\\
It is worth mentioning that using the LW representation is straightforward to generate different vector beams. It allows to study fundamentals properties for electromagnetic beams  \cite{IrvingInvBeam}, the transversal component of the Poynting vector
\cite{IrvingPvec} and the orbital angular momentum for nonparaxial beams \cite{IrvingOApp} among others problems. The following section shows how an extended optical theorem using these results.

\begin{figure} 
	\centering
	\includegraphics[scale=0.5]{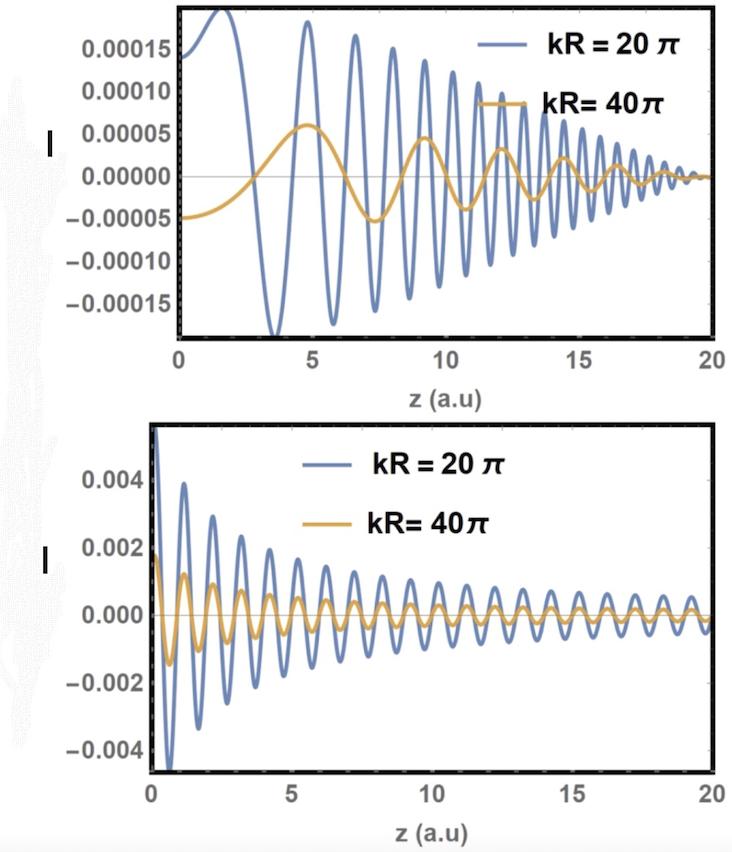}\quad 
	\caption{ Upper: the $sinc$  beam  given by Eq. \ref{eq13}. Bottom: the Gaussian beam Eq. \ref{eq14} using $n=3$. The functions are labeled in terms of $kR$. In both cases, the energy flux is integrated over a circular aperture of radius $D$. This oscillatory behavior of the intensity between negative and positive amplitudes has been previously reported. (See Ref. \cite{Mishchenko3} and references therein).} 
	\label{Fig1}
\end{figure}

\newpage
\section{Results and Methods: }
\subsection{the optical cross section using a longitudinal  wave spectrum}
Using the LW wave spectrum represented  by Eq. \ref{LongInvB}  and substituting into Eq \ref{eq2}, we put forward  a general optical theorem using LW for arbitrary beam using a  longitudinal spectrum as

\begin{eqnarray}
		\label{GenLonSz0}
	&I \approx  \int_{S} \varphi ^* \varphi \,\, da \\
& + \frac{2\pi }{R}
	\mathbf{Re}  \int_S \left(  f^*    \int_{-\omega/c}^{\omega/c}  J_0\left( \rho \sqrt{ \omega^2/c^2-k_z^2}\right) S(k_z) \exp{ \left( ik_z z + ik \frac{\rho^2}{2R} \right)} dk_z   \right) da \nonumber, 
\end{eqnarray}
It is well know for the OT using a plane wave, the first integral is related to the absorption \cite{RNewton,Jackson,Bohren}, it can be denoted as  $A(s) = \int_{S} \varphi ^* \varphi \,\, da$,  the  second expression is related  with extinction or the optical theorem \cite{Markel,Mishchenko1,Mishchenko2,Mishchenko3} .  After  an integration, due to the cylindrical symmetry  in the interval $0 \leq  \theta \leq 2 \pi$, we obtain 

\begin{eqnarray}
\label{GenLonSz}
&I  \approx A(s) \\
&+   \frac{4\pi }{R}
\mathbf{Re}\left(  f^*    \int_{0}^{D}  \int_{-\omega/c}^{\omega/c}  J_0\left( \rho \sqrt{ \omega^2/c^2-k_z^2}\right) S(k_z)  \exp{ \left( ik_z z + ik \frac{\rho^2}{2R} \right)} dk_z   d\rho \right) \nonumber  \nonumber.	
\end{eqnarray}
After the evaluation, the first term after a well-defined region is a constant value, and the second integral term is related to the extinction, precisely solvable.
It is worth here to remark that these results  recover the OT plane wave case: when $\rho \rightarrow 0$, $J_0 (0) = 1$ and $z=0$,  $S(k_z)$  takes a constant value.
\\
In the following section, we show analytical solutions for Eq. \ref{GenLonSz}  using two different probe beams. It means using two longitudinal spectrum $S(k_z)$, a  cylindrical beam using  Eq. \ref{Eq9_Sz0} and a Gaussian beam using Eq. \ref{GenLonSz0}. 

\subsection{Example 1: Optical cross section  for constant longitudinal spectrum}
\noindent
Using a  cylindrical $sinc$ beam Eq. \ref{Eq9_Sz0}  into Eq. \ref{GenLonSz}. It  allows to obtain an optical cross-section solution as

\begin{eqnarray}
	\label{eq13}	\nonumber
I \approx & \,\, 
\mathbf{Re}
\Biggl\{ 
\frac{\pi \sqrt{\pi }  }{k} \frac{\gamma f^{*}}{\sqrt{k R}}  \exp{ \left[ -i k \left(R^2+z^2\right)/2 R \right]} \times
\Bigg(
\mathbf{erfi}\left[ \frac{\gamma}{\sqrt{k R}}  (R-z)\right] \\
&+\mathbf{erfi}\left[ \frac{\gamma}{\sqrt{ k R}} (R+z)\right]
 -\mathbf{erfi}\left[\frac{\gamma}{\sqrt{k R}} \left(R-\sqrt{D^2+z^2}\right)\right] \nonumber \\ 
 &-\mathbf{erfi}
\left[ \frac{\gamma}{\sqrt{k R}} \left(R+\sqrt{D^2+z^2}\right)\right]
\Bigg) \Biggr\},		
\end{eqnarray}
where $\gamma= (1+i)/2$ and  $k=\omega/c$ is  the wave vector. The analytical results are given in measurable parameters such as  $D$ ( detector diameter's), the propagation distance $z$, and the scaled product of $kR$. (See Refs \cite{Mishchenko1,Mishchenko2,Mishchenko3,Markel}
For the plane wave case). In Fig \ref{Fig1}, we show the numerical behavior when the size $D$ detector is fixed, and $z$ varies is shown. The extinction amplitude decreases to zero as expected, while the intensity directly vanishes when $D \rightarrow 0$ (point particle).
\subsection{Example 2: Optical cross section for a Gaussian beam}
\noindent
Substituting Eq. \ref{Eq_10aGSinc}, into  Eq. \ref{eq2} the optical cross section for a longitudinal Gaussian beam is  given by

\begin{eqnarray}
		\label{eq14} \nonumber
		I \approx&  \,\, \mathbf{Re} \sum_{n=0}^{\infty} 
		\Biggl\{ - \frac{ 2 \pi\sqrt{\pi }  }{ k} \frac{\gamma f^{*} }{\sqrt{k R}} \exp{ \left[ -\frac{i \bar{k}_z n \pi }{2 \Lambda}-\frac{\pi ^2 n^2}{b^2 \Lambda^2}-\frac{i \left(  k^2 R^2 + \left(n \pi + k z\right)^2 \right)  }{2 k R} \right]}\times
		\\  
		&\Bigg( \mathbf{erfi} \left[\frac{  \gamma \left(k R-\sqrt{(k z+\pi n)^2}\right)}{\sqrt{k R}}\right]
		- \mathbf{erfi}\left[ \frac{ \gamma \left(k R+\sqrt{(k z+ \pi n)^2}\right)}{\sqrt{k R}} \right]\\
		&+ \mathbf{erfi} \left[ \frac{  \gamma \left( k R+\sqrt{D^2 k^2+(k z+ \pi n)^2}\right)  }{\sqrt{k R}} \right] 
		+ \mathbf{erfi} \left[ \frac{ \gamma 
			\left(k R-\sqrt{D^2 k^2+
				(k z+\pi n)^2}\right)}{\sqrt{k R}} \right]\Bigg)
		\Biggr\},
		\nonumber
\end{eqnarray}
Fig \ref{Fig1} shows the numerical behavior in the nonparaxial regime \cite{Roger} using the following values: $\bar{k}_z = 0.4 \omega/c$, $\Lambda= 2 \omega/c$ and $b = 4/\bar{k}_z$  where $\omega=c=1$.
After several simulations varying $n$ for small values, the series converges, the cross-section has oscillatory behavior, and the amplitude decreases. These results are in agreement with the case of a circular detector reported, which showed fast and decaying oscillations in time \cite{Markel,Mishchenko3}. In this work, only a circular sensor and fixed-size detector were considered. Additional analysis such as the numerical convergence would be interesting to include the effects of interference terms \cite{Markel}. 
It may be of interest to make further numerical simulations and comparisons with several reported publications, also
to explore using different cylindrical beams, but it is outside the scope of this work,  is open for future research, and can be addressed elsewhere.

\section{Summary}
We presented a generalized optical theorem using  Localized Waves. We show an analytical solution for arbitrary LW beams for the scattering cross-section problem using this representation. Our findings suggest that (Eq. \ref{GenLonSz0}) can be applied in optics and acoustics beams scattering problems.
We show  general
solutions for the optical theorem for two cases, a cylindrical beam (Eq. \ref{eq13}) and a longitudinal Gaussian beam (Eq. \ref{eq14}).
This new heuristic method can work under the appropriate $S(k_z)$ under well-defined physical initial conditions and experimental parameters.   Another application is to measure the cross-section in optical/acoustic problems, for example, to measure the scattering experimentally using  "Frozen waves" \cite{Zamboni,Joel} or another LW. Since the parameters presented can be connected.
Also, considering the vector effect   \cite{Krasavin} in the extinction. 
Several interesting acoustic scattering problems 
as the spin propagation  \cite{Yang},  orbital angular momentum  \cite{KBliok1},  vortex modes in synthetic fluids \cite{IrvingIBS}, evanescent Bessel waves \cite{IrvingKIAS} could be analyzed using LW. This method can suit to study experimentally new structured beams, such as carving beams recently reported \cite{Michel2021}.

\section{Acknowledgments}
The author acknowledged 	financial   support by Basic Science Research Program through the National Research Foundation of Korea (NRF) funded by the Ministry of  Science and ICT [NRF-2017R1E1A1A01077717].

\end{document}